# Systematic method to evaluate energy dissipation in adiabatic quantum-flux-parametron logic

Taiki Yamae,[1,*] Naoki Takeuchi,[2] and Nobuyuki Yoshikawa[1,2]

[1] Department of Electrical and Computer Engineering, Yokohama National University, 79-5 Tokiwadai, Hodogaya, Yokohama 240-8501, Japan

[2] Institute of Advanced Sciences, Yokohama National University, 79-5 Tokiwadai, Hodogaya, Yokohama 240-8501, Japan

[*] yamae-taiki-yw@ynu.jp

**Abstract.** Adiabatic quantum-flux-parametron (AQFP) logic is an energy-efficient superconductor logic. It operates with zero static power dissipation and very low dynamic power dissipation owing to adiabatic switching. In previous numerical studies, we have evaluated the energy dissipation of basic AQFP logic gates and demonstrated sub-$k_\mathrm{B}T$ switching energy, where $k_\mathrm{B}$ is the Boltzmann's constant and $T$ is the temperature, by integrating the product of the excitation current and voltage associated with the gates over time. However, this method is not applicable to complex logic gates, especially those in which the number of inputs is different from the number of outputs. In the present study, we establish a systematic method to evaluate the energy dissipation of general AQFP logic gates. In the proposed method, the energy dissipation is calculated by subtracting the energy dissipation of the peripheral circuits from that of the entire circuit. In this way, the energy change due to the interaction between gates, which makes it difficult to evaluate the energy dissipation, can be deducted. We evaluate the energy dissipation of a majority gate using this method.

I. Introduction

Various superconductor logic families have been developed to realize future energy-efficient computing systems. Rapid single-flux-quantum (RSFQ) logic[1] is one of the most well-known superconductor logic families and several prototype RSFQ microprocessors have been demonstrated[2]. In RSFQ logic, the power dissipation is dominated by the static power dissipation caused by bias resistors. The power dissipation of an RSFQ circuit is given by $<I_b>V_bN$, where $<I_b>$ is the average bias current, $V_b$ is the bias voltage, and $N$ is the number of Josephson junctions. To eliminate static power dissipation in the bias network, more energy-efficient superconductor logic families, such as energy-efficient RSFQ (ERSFQ) logic[3] and reciprocal-quantum-logic (RQL)[4], have been proposed. With the static power dissipation eliminated, the power dissipation in these energy-efficient logic families is dominated by the dynamic energy dissipation of Josephson junctions, given by $\alpha I_c\Phi_0$, where $I_c$ is the critical current of Josephson junctions, $\Phi_0$ is the flux quantum, and $\alpha$ is a logic-dependent factor. $\alpha$ is 0.75 and 0.33 for ERSFQ logic and RQL, respectively[3,4]. The power dissipation of a circuit designed using ERSFQ or RQL is thus given by $\alpha<I_c>\Phi_0Nf$, where $<I_c>$ is the average critical current and $f$ is the operating frequency. The literature regarding RQL[5] has experimentally verified this energy estimation method and reported a power dissipation of 0.626 nW per junction at 6.21-GHz operation, which is much smaller than that in RSFQ logic (approximately 300 nW per junction). As described above, energy estimation methods have been established for both RSFQ and more energy-efficient logic families, which are important for estimating the energy efficiency of large-scale superconductor circuits[6,7].

The adiabatic quantum-flux-parametron (AQFP)[8] is an adiabatic superconductor logic family based on the quantum-flux-parametron[9,10]. The switching energy (energy dissipation per clock cycle) of an AQFP gate can be reduced to less than $I_c\Phi_0$ by using adiabatic switching[11,12], in which the potential energy of a logic gate gradually changes from a single-well shape to a double-well shape so that the logic state can change quasi-statically. In a previous study[13], we calculated the switching energy of an AQFP buffer and numerically demonstrated a switching energy less than the thermal energy $k_BT$, where $k_B$ is Boltzmann's constant and $T$ is the temperature. We also calculated the energy dissipation of reversible AQFP logic gates[14]. In both the above, the energy dissipation was calculated by integrating over time the excitation current and voltage associated with the gate under test. However, this method is not applicable to more complex logic gates, especially those in which the number of inputs is different from the number of outputs. This is because in AQFP logic, not only do the power supplies perform work on the logic gates, but the logic gates also perform work on the power supplies, whereas in conventional superconductor logic it is only the power supplies that perform work. Thus, the work performed by the power supplies is not always equal to the energy dissipation, which makes it difficult to calculate the energy dissipation of AQFP logic gates, as will be shown later. Also, note that the energy dissipation of some AQFP logic gates strongly depends on the input data[14]. Thus, unlike in conventional superconductor logic, the energy dissipation of an AQFP logic gate cannot be simply determined from the junction count. As a result, it is currently difficult to estimate the energy dissipation of complex AQFP circuits.

In the present study, we propose a systematic method to evaluate the energy dissipation of arbitrary AQFP logic gates by closely investigating the work, energy, and dissipation in AQFP logic. First, we explain how the energy dissipation of a buffer can be calculated and why it is difficult to calculate the energy dissipation of more complex logic gates. Then, we describe the proposed method, where the energy dissipation of the logic gate under test is calculated by subtracting the energy dissipation of the peripheral circuits from that of the entire circuit. Using this method, we calculate the energy dissipation of a somewhat complex logic gate, a majority (MAJ) gate.

## II. Energy evaluation of AQFP buffers

First, we consider the energy dissipation of the simplest AQFP logic gate, i.e., a buffer. To calculate the energy dissipation of an AQFP buffer, we prepare a somewhat long buffer chain, as shown in Fig. 1(a). This is because the logic gates near the input current sources and those near the terminating ports behave differently from the other gates[15]. Each buffer is excited by the ac excitation current $I_{x,i}$ and operates as follows. Let $\phi_1$ and $\phi_2$ be the phase differences across $J_1$ and $J_2$, respectively. As $I_{x,i}$ increases, both $\phi_1$ and $\phi_2$ increase from 0. The logic state of the buffer is determined by which of $\phi_1$ and $\phi_2$ reaches $\pi$ first. If there is no input current, i.e., $I_{in,i} = 0$, then $\phi_1$ and $\phi_2$ increase at the same velocity, so the logic state is stochastically determined. This indicates that both the logic 0 and 1 states are stable, i.e., the potential energy shape is symmetrical. On the other hand, if a positive (negative) input current is applied, i.e., $I_{in,i} > 0$ ($I_{in,i} < 0$), then $\phi_1$ ($\phi_2$) reaches $\pi$ first because a part of $I_{in,i}$ and the screening current due to the magnetic flux applied by $I_{x,i}$ sum up in the same polarity at $J_1$ ($J_2$). This indicates that the logic 1 (0) state is stable and has lower energy than the logic 0 (1) state, i.e., the potential energy shape is tilted towards a logic 1 (0). The logic state of the buffer is distinguished by the polarity of the state current $I_{st,i}$, which appears through the load inductor $L_q$. A positive $I_{st,i}$ represents a logic 1, while a negative $I_{st,i}$ represents a logic 0. The output current $I_{out,i}$ is generated through the transformer composed of $L_q$ and the output inductor $L_{out}$. Figure 1(b) illustrates waveforms showing how the data propagates from buffer $i$-1 to buffer $i$+1, where $i \in \{1, 2, 3, …\}$. $I_{x,i-1}$, $I_{x,i}$, and $I_{x,i+1}$ are the excitation currents applied to buffers $i$-1, $i$, and $i$+1, respectively. $I_{st,i-1}$, $I_{st,i}$, and $I_{st,i+1}$ are the state currents of buffers $i$-1, $i$, and $i$+1, respectively. For simplicity, we assumed the profile of the excitation currents to be trapezoidal, whereas, in the actual circuit implementation, sinusoidal excitation currents are used[16]. This figure shows that the data, a logic 1, propagates as the buffers are excited in turn by the excitation currents.

We now discuss how to calculate the energy dissipation of buffer $i$ using the time evolution of the potential energy of the buffer, shown in Fig. 1(c). A through E represent the stages during excitation defined in Fig. 1(b). At stage A, $I_{x,i}$ is in the low level, so the potential energy of buffer $i$ is a single-well shape. Between stages A and B, buffer $i$ switches to a logic 1 as $I_{x,i}$ ramps up gradually, because buffer $i$-1 applies a positive $I_{in,i}$ to buffer $i$ and tilts its potential energy towards a logic 1. During this switching event, the potential energy evolves from a single-well shape into a double-well shape. As a result, the energy of buffer $i$ changes by $\Delta E_{exc,i}$ (> 0) at stage B. Between stages B and C, $I_{x,i-1}$ ramps down and buffer $i$-1 is reset. At stage C, buffer $i$-1 does not apply a signal current to buffer $i$ and thus the potential energy of buffer $i$ is no longer tilted. As a result, the energy of buffer $i$ changes by $\Delta E_{fwd,i-1,i}$ (> 0) at stage C. Between stages C and D, $I_{x,i+1}$ ramps up and buffer $i$+1 switches to a logic 1. Since AQFP gates are symmetrical from the viewpoint of the schematics, buffer $i$+1 applies a back-action current to buffer $i$ [14]. As a result, buffer $i$ is tilted again towards a logic 1 and its energy changes by $\Delta E_{bwd,i+1,i}$ (< 0) at stage D. Finally, $I_{x,i}$ ramps down between stages D and E and buffer $i$ is reset. Thus, the energy of buffer $i$ changes by $\Delta E_{res,i}$ (< 0) at stage E. From the above, the energy change of buffer $i$ during an excitation cycle (from stage A to E) is given by:

$$\Delta E_i = \Delta E_{\text{exc},i} + \Delta E_{\text{res},i} + \Delta E_{\text{fwd},i-1,i} + \Delta E_{\text{bwd},i+1,i}, \qquad (1)$$

where $\Delta E_{exc,i}$ and $\Delta E_{res,i}$ are the energy changes caused by the work performed by $I_{x,i}$, while $\Delta E_{fwd,i-1,i}$ and $\Delta E_{bwd,i+1,i}$ are the energy changes caused by the work performed by $I_{x,i-1}$ and $I_{x,i+1}$, respectively. This equation indicates that $I_{x,i}$ performs work not only on buffer $i$ but also on buffers $i$-1 and $i$+1. Therefore, since the work is given by the sum of the free energy change and dissipated work[17], the work performed by $I_{x,i}$ can be given by:

$$W_i = \Delta E_{\text{exc},i} + \Delta E_{\text{res},i} + \Delta E_{\text{bwd},i,i-1} + \Delta E_{\text{fwd},i,i+1} + E_{\text{diss},i}, \tag{2}$$

where $E_{\text{diss},i}$ is the dissipated work, which is defined as the energy dissipation of buffer $i$. In the present study, we assume that every logic operation is deterministic (i.e., there is no entropy change), so we do not distinguish between free energy change and energy change. Thus, Eq. 2 is described in the form of energy change, rather than free energy change. We assume that the buffer chain is symmetrical, i.e., $L_q = L_{\text{out}}$. Therefore, when buffer $i$ is excited, the back-action current whose amplitude is the same as $I_{\text{st},i}$ appears through the input port, i.e., $I_{\text{st},i} = -I_{\text{in},i}$. This back-action current $(-I_{\text{in},i})$ flows to buffer $i$-1 and applies the magnetic flux of $-k_{\text{out}}(L_q L_{\text{out}})^{0.5} I_{\text{in},i}$ to buffer $i$-1, which is the same as the flux applied to buffer $i$+1, $k_{\text{out}}(L_q L_{\text{out}})^{0.5} I_{\text{st},i}$, because $I_{\text{st},i} = -I_{\text{in},i}$. Thus, buffer $i$ applies the same magnetic fluxes to buffers $i$-1 and $i$+1, so that it is reasonable to assume that the forward work performed by $I_{\text{x},i}$ on buffer $i$+1 and the backward work performed by $I_{\text{x},i}$ on buffer $i$-1 cancel each other out, that is:

$$\Delta E = \Delta E_{\text{fwd},i,i+1} = -\Delta E_{\text{bwd},i,i-1}, \tag{3.a}$$

$$\Delta E_{\text{exc},i} = -\Delta E_{\text{res},i}. \tag{3.b}$$

Using these equations, Eq. 2 can be reduced to $W_i = E_{\text{diss},i}$, which shows that the energy dissipation of buffer $i$ can be obtained from the work performed by $I_{\text{x},i}$. The work performed by $I_{\text{x},i}$ can be calculated by integrating the product of $I_{\text{x},i}$ and the excitation voltage across the excitation inductors [$L_{\text{x}1}$ and $L_{\text{x}2}$ in Fig. 1(a)] over time. In previous studies[13,15], we calculated the energy dissipation of buffers using this method and obtained reasonable results compared to analytical estimations.

**III. Energy evaluation of general AQFP logic gates**

Next, we consider the general logic gate shown in Fig. 2, where the circuit under test (CUT) is surrounded by peripheral buffers to isolate it from the input current sources and terminating ports, as with the case of a buffer chain. The CUT is powered by the excitation current $I_{x,i}$ and has $m$ input ports and $n$ output ports, where $m$ and $n$ are arbitrary natural numbers. In a similar way to the derivation of Eq. 2, the work performed by $I_{x,i}$ can be given by:

$$W_{\text{CUT}} = \Delta E_{\text{exc},i} + \Delta E_{\text{res},i} + \sum_{j=i-m}^{i-1} \Delta E_{\text{bwd},i,j} + \sum_{j=i+1}^{i+n} \Delta E_{\text{fwd},i,j} + E_{\text{diss,CUT}}, \quad (4)$$

where $E_{\text{diss,CUT}}$ is the energy dissipation of the CUT. For simplicity, we assume that Eq. 3 is true for the CUT, and thus Eq. 4 is reduced to:

$$W_{\text{CUT}} = E_{\text{diss,CUT}} - (m - n)\Delta E, \quad (5)$$

which clearly shows that if $n \neq m$ (e.g., the CUT is an MAJ gate or an AND gate), the energy dissipation of the CUT cannot be obtained from the work performed by $I_{x,i}$, unlike the case of a buffer. Therefore, complex logic gates in which the number of inputs is different from the number of outputs require different methods for energy evaluation.

We now look at the total work performed on the entire circuit shown in Fig. 2. The total work $W_{\text{tot}}$ performed by the excitation currents $I_{x,1}$ through $I_{x,N}$, where $N$ is the total number of gates, is given by:

$$W_{\text{tot}} = \sum_{j=1}^{N} W_j = \sum_{j=1}^{N} E_{\text{diss},j}. \quad (6)$$

This equation does not include terms representing the energy change as the total energy of the entire circuit shown in Fig. 2 does not increase or decrease for an excitation cycle. Eq. 6 shows that while the energy dissipation of the CUT cannot be directly obtained from the work, the total energy dissipation of the entire circuit can be obtained from the total work, and that the energy dissipation of the CUT can be obtained by subtracting the energy dissipation of the peripheral buffers from that of the entire circuit. The work $W_{\text{per}}$ performed by the excitation currents ($I_{x,1}$ through $I_{x,i-1}$ and $I_{x,i+1}$ through $I_{x,N}$) coupled to the peripheral buffers is given by:

$$W_{\text{per}} = W_{\text{tot}} - W_{\text{CUT}}$$

$$= \sum_{j=1}^{i-1} E_{\text{diss},j} + \sum_{j=i+1}^{N} E_{\text{diss},j} + (m-n)\Delta E \quad (7)$$

$$= E_{\text{diss,per}} + \Delta E_{\text{per}},$$

where

$$E_{\text{diss,per}} = \sum_{j=1}^{i-1} E_{\text{diss},j} + \sum_{j=i+1}^{N} E_{\text{diss},j}, \quad (8.\text{a})$$

$$\Delta E_{\text{per}} = (m-n)\Delta E. \quad (8.\text{b})$$

Importantly, $E_{\text{diss,per}}$ and $\Delta E_{\text{per}}$ in Eq. 7 can be individually obtained by simulating $W_{\text{per}}$ as a function of the operating frequency $f$. This is because buffers always operate adiabatically[13], so that $E_{\text{diss,per}}$ should be proportional to $f$, while $\Delta E_{\text{per}}$ is independent of $f$. Therefore, $\Delta E_{\text{per}}$ can be specified by extrapolating the simulation results of $W_{\text{per}}$ as a function of $f$, as will be shown later.

From the above, the energy dissipation of the CUT is given by:

$$E_{\text{diss,CUT}} = W_{\text{tot}} - W_{\text{per}} + \Delta E_{\text{per}}. \quad (9)$$

Note that $W_{\text{tot}}$ and $W_{\text{per}}$ can be obtained by integrating the excitation currents and voltages of the associated logic gates over time, and that $\Delta E_{\text{per}}$ can be obtained by extrapolating the simulation results of $W_{\text{per}}$, even though we cannot obtain $\Delta E_{\text{per}}$ analytically.

**IV. Numerical simulation**

Here, we take an MAJ gate as an example to show how to calculate the energy dissipation of general logic gates using Eq. 9. We use the schematic shown in Fig. 3(a), where a three-input MAJ gate is surrounded by peripheral buffers. The MAJ gate is composed of three buffers, the output currents of which are merged to conduct majority logic[18]. The entire circuit is powered and clocked by a pair of sinusoidal excitation currents with a phase separation of 90° ($I_{x1}$ and $I_{x2}$), which apply an ac magnetic flux with an amplitude of $0.5\Phi_0$ to each gate. The dc offset current $I_d$ applies a dc magnetic flux of $0.5\Phi_0$ to each gate. In this way, logic operations are performed with a phase separation of 90°. More details regarding excitation methods in AQFP logic can be found in the literature[16]. It is noteworthy that the difference in the profile of excitation currents does not make difference on the above discussions and equations because none of the equations (Eqs. 1 through 9) include the profile of excitation currents. In the numerical simulation, we use AQFP logic gates based on the previous study[16] and device parameters for the AIST 10-kA/cm² Nb high-speed standard process (HSTP)[16]. The numerical simulation is conducted using the Josephson circuit simulator, JSIM[19]. Figure 3(b) shows example transient analysis results for the MAJ gate for $f$ = 5 GHz, where $f$ corresponds to the frequency of $I_{x1}$ and $I_{x2}$. $I_{ina}$, $I_{inb}$, and $I_{inc}$ are the input currents. $I_{sta}$, $I_{stb}$, and $I_{stc}$ are the state currents of buffers A, B, and C [see Fig. 3(a)], respectively. $I_{stq}$ is the state current of buffer Q. The figure shows that the MAJ gate operates correctly. Note that the latency between $I_{stq}$ and $I_{sta}$ through $I_{stc}$ is two excitation cycles.

First, $W_{tot}$ is calculated as follows:

$$W_{tot} = \int_0^{1/f} (I_{x1}V_{x1} + I_{x2}V_{x2})dt, \qquad (10)$$

where $V_{x1}$ and $V_{x2}$ are the voltages across the current sources $I_{x1}$ and $I_{x2}$, respectively [see Fig. 3(a)]. Figure 4(a) shows the calculated $W_{tot}$ as a function of $f$ for $(a, b, c)$ = (1, 1, 1) and (1, 0, 1), where $a$, $b$, and $c$ are the logic of the input currents. Note that $W_{tot}$ is always positive since $W_{tot}$ represents the energy dissipation of the entire circuit, as shown in Eq. 6. The reason why $W_{tot}$ varies depending on the inputs is because the thermodynamically irreversible operation causes a non-adiabatic state transition[14]. For $(a, b, c)$ = (1, 0, 1), the inputs cannot be restored from the output ($q$ = 1), i.e., the MAJ gate cannot operate thermodynamically reversibly and causes a non-adiabatic state transition. On the other hand, the MAJ gate can operate thermodynamically reversibly for $(a, b, c) = (1, 1, 1)$ because $q = 1$ restores $(a, b, c) = (1, 1, 1)$ by operating the MAJ gate in a time-reversed way. The details of reversibility and energy dissipation in AQFP logic can be found in the literature[14]. Then, $W_{CUT}$ is calculated as follows:

$$W_{CUT} = \int_0^{1/f} I_{x1} V_{CUT} dt, \tag{11}$$

where $V_{CUT}$ is the voltage across the excitation inductors of the MAJ gate [see Fig. 3(a)]. Note that Eq. 11 includes only $I_{x1}$ because $I_{x2}$ is not coupled to the MAJ gate. Also, $I_d$ is dc and thus does not perform work. Figure 4(b) shows the calculated $-W_{CUT}$ as a function of $f$. Unlike $W_{tot}$, $W_{CUT}$ is negative because $W_{CUT}$ includes the energy change due to interaction between gates (see Eq. 5), which can be negative depending on the values of $m$ and $n$. This clearly shows that the energy dissipation of the MAJ gate cannot be directly obtained from $W_{CUT}$, unlike the case of a buffer. Next, $W_{per}$ is calculated as $W_{per} = W_{tot} - W_{CUT}$, which is shown in Fig. 4(c). For both $(a, b, c) = (1, 1, 1)$ and $(1, 0, 1)$, $W_{per} = E_{diss,per} + \Delta E_{per}$ (see Eq. 7) approaches non-zero values as $f$ decreases. This is because $E_{diss,per}$ (which represents the energy dissipation of buffers) is proportional to $f$,[13] and $\Delta E_{per}$ (which represents the energy change owing to the gate interaction) is independent of $f$. Therefore, $\Delta E_{per}$ can be obtained by extrapolating the simulation results of $W_{per}$ as a function of $f$ with the linear regression $W_{per} = \alpha f + \beta$, where $\alpha$ (= $E_{diss,per}/f$) and $\beta$ (= $\Delta E_{per}$) are the fitting parameters. The dashed lines in Fig. 4(c) are the linear regression for the simulation results, which shows that $\Delta E_{per} = 1.48 \times 10^{-20}$ J and $2.57 \times 10^{-20}$ J for $(a, b, c) = (1, 1, 1)$ and $(1, 0, 1)$, respectively. Finally, $E_{diss,CUT}$ is calculated using Eq. 9 with $W_{tot}$, $W_{per}$, and $\Delta E_{per}$ obtained above, as shown in Fig. 4(d). As mentioned above, $E_{diss,CUT}$ depends on the combination of $(a, b, c)$; $E_{diss,CUT}$ approaches a non-zero value for $(a, b, c) = (1, 0, 1)$ as $f$ decreases due to the non-adiabatic state transition, whereas $E_{diss,CUT}$ decreases in proportion to $f$ for $(a, b, c) = (1, 1, 1)$. Fig. 4(d) demonstrates that the proposed method can evaluate the energy dissipation of complex AQFP logic gates.

## V. Conclusion

We established a systematic method to evaluate the energy dissipation of complex AQFP logic gates, especially those in which the number of inputs is different from the number of outputs. The energy dissipation is calculated by subtracting the energy dissipation of the peripheral buffers from that of the entire circuit. We calculated the energy dissipation of an MAJ gate using this method. The next step is to evaluate the energy dissipation of other AQFP logic gates, such as an AND gate and a full adder, using the proposed method.

**Acknowledgements**

The present study was supported by JSPS KAKENHI (Grants No. 18H01493 and No. 26220904).

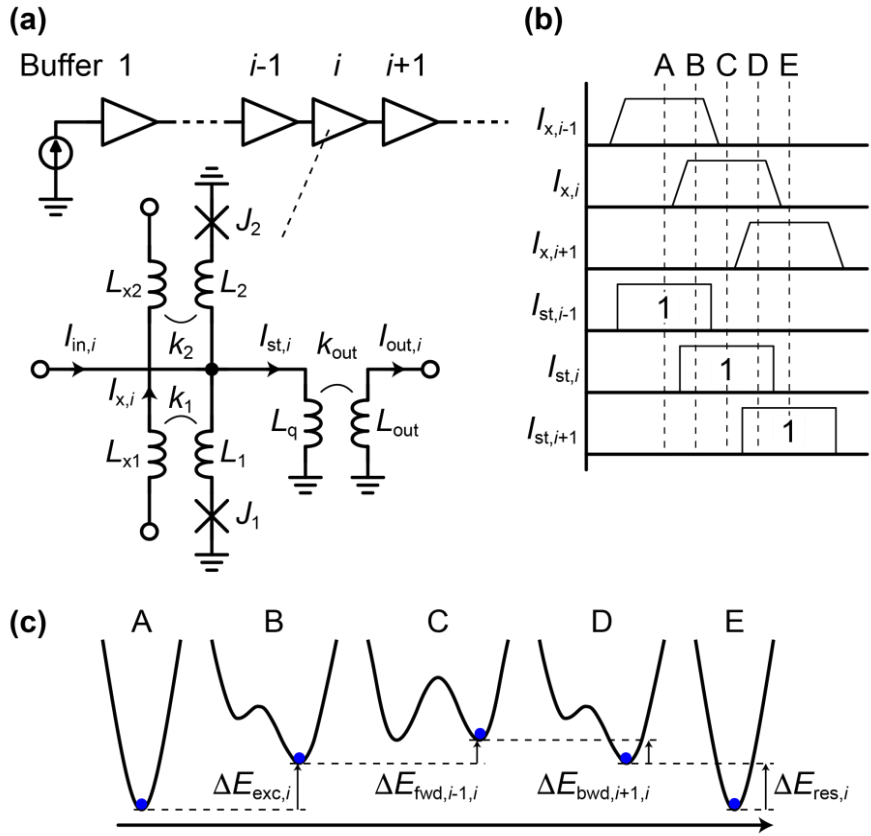

Fig. 1  Buffer chain. (a) Schematic. (b) Waveform. A logic 1 propagates through buffers $i$-1, $i$, and $i$+1. (c) Time evolution of the potential energy of buffer $i$. The potential is tilted by buffers $i$-1 and $i$+1.

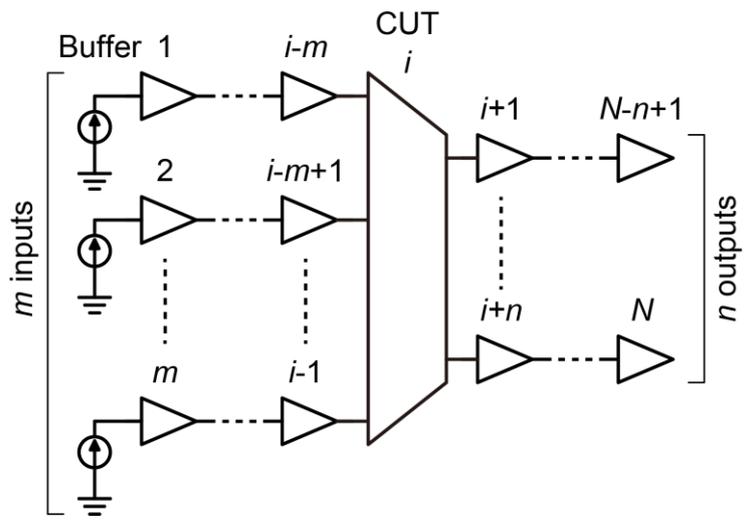

Fig. 2    General logic gate with *m* inputs and *n* outputs.

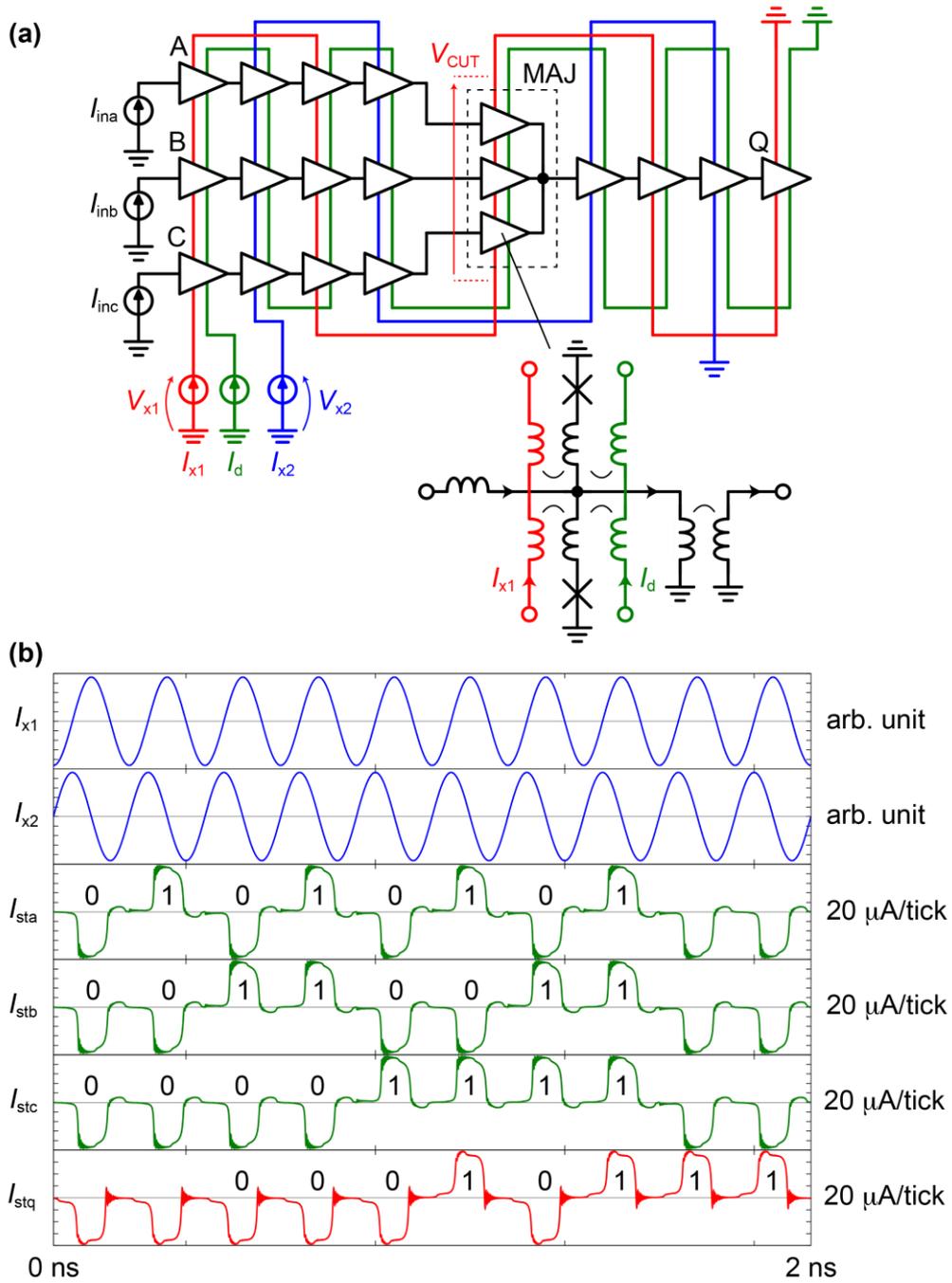

Fig. 3　MAJ gate. (a) Schematic for simulation. The MAJ gate is isolated from the input current sources and terminating ports by the peripheral buffers. (b) Transient analysis results at 5 GHz.

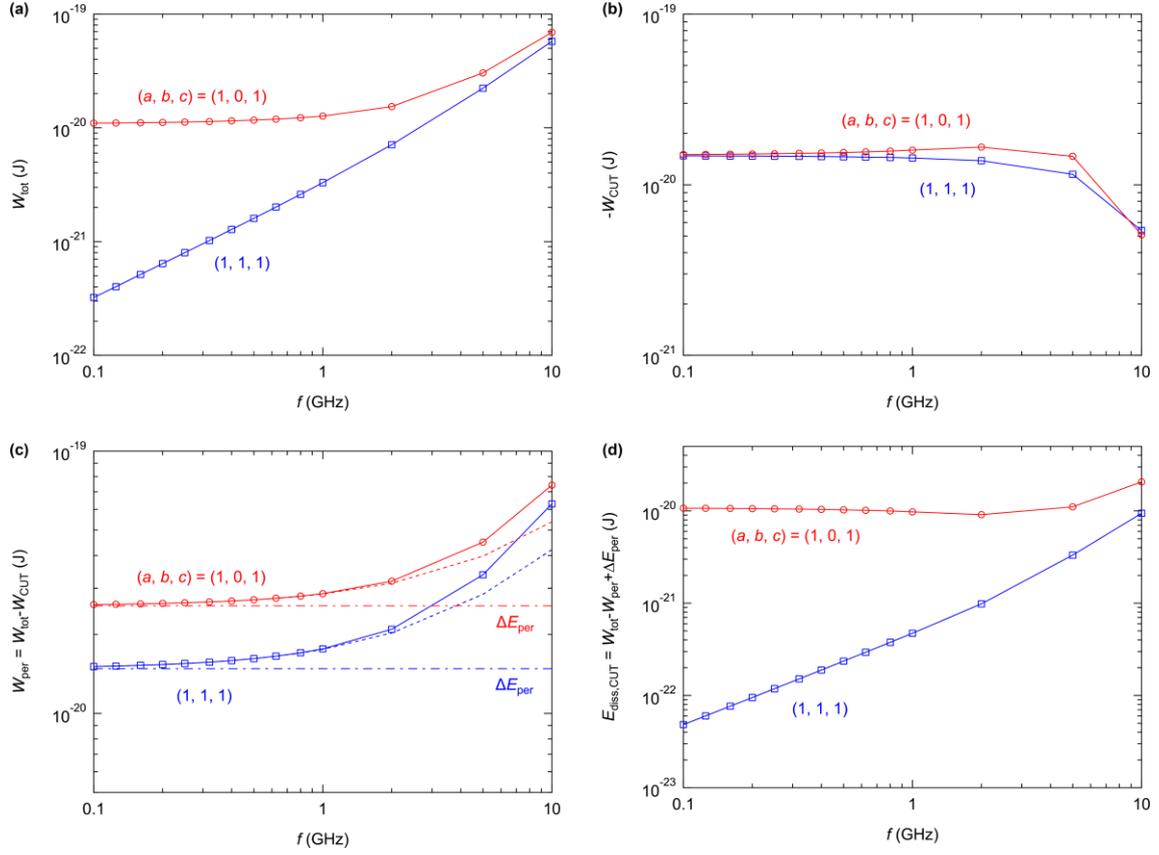

Fig. 4  Work and dissipation associated with an MAJ gate. (a) Total work performed by $I_{x1}$ and $I_{x2}$. (b) Work performed by the excitation current coupled to the MAJ gate. (c) Work performed by the excitation currents coupled to the peripheral buffers. (d) Energy dissipation of the MAJ gate.